# Surface Wetting Study via Pseudocontinuum Modeling

*Meysam Makaremi[1], Myung S. Jhon[1,\*], Meagan S. Mauter[2,3], and Lorenz T. Biegler[1]*

[1]Department of Chemical Engineering, [2]Department of Civil and Environmental Engineering, and [3]Department of Engineering & Public Policy, Carnegie Mellon University, Pittsburgh, Pennsylvania 15213.

ABSTRACT Accurate estimation of surface wettability for various degrees of hydrophobicity becomes increasingly important in the molecular design of membrane. In this paper, we develop simple yet physically realistic model for estimating contact angle via hybridizing molecular dynamics and pseudocontinuum theory. Molecular dynamics simulations were carried out to compute the macro-scale contact angle between a water droplet and smooth walls from the nanoscale calculations. A macro-level droplet including countless degrees of freedom due to an infinite number of molecules is impossible to be studied directly via atomistic simulations. To resolve this issue, we employed two approaches consisting of the pseudocontinuum approximation and the modified Young-Laplace equation. The former involves the 9-3 Lennard-Jones (L-J) potential and can drastically reduce the degrees of freedom in molecular simulations, while the latter relates the mesoscale contact angle to the realistic one. We altered different parameters including the liquid-surface potential characteristics and the temperature, and calculated the water contact angle by leveraging the mass density profile fitting method to predict the broad spectrum of hydrophobic and hydrophilic substrates. The computational results were compared with the experimental data for various materials including graphite, silicon, and metals. This study suggests that pseudocontinuum modeling is an accurate approach to probe surface wettability for various processes at a low computational cost.

1. Introduction

Wetting tendency is a determining variable in many physical and chemical processes such as water treatment, oil recovery, surface coating, adhesion, and lubrication.[1-7] For example, in reverse osmosis filtration and coating where bonding is essential, this variable is a key feature, while, in membrane distillation and biological processes where water repellence or self-cleaning is required, it is a limiting issue.[8-10] Affected by surface properties, wetting tendency is the result of a competition between liquid cohesive forces and liquid-surface adhesive forces, which can be analyzed by measuring the liquid contact angle on the surface or the surface energy. Therefore it is desirable to develop simple yet physically realistic model that is descriptive for surface energies from atomistic/molecular level approaches to design macroscopic performance of membrane system. When a liquid droplet in contact with a surface making a contact angle smaller than 90° illustrates the non-wetting tendency (hydrophobicity), whereas if the liquid is extensively adsorbed by the surface and develops a contact angle greater than 90°, it presents wetting tendency (hydrophilicity).[11, 12]

The contact angle has been the subject of numerous experimental studies.[13-16] Bernardin et al.[17] examined the effect of temperature and pressure on the water contact angle of aluminum surfaces, and found that the temperature plays a significant role on wettability at high temperatures. Öner et al.[18] investigated the wetting properties of silicon surfaces for various types of posts; such as square, rhombus and star bumps; formed by photolithography and hydrophobized by silanization chemistry. Ranella et al.[19] illustrated Si micro- and nano-rough patterns prepared by the femtosecond laser structuring method can show different wetting tendencies stemming out from the roughness size controlled by the laser pulse fluence. Grundke et al.[20] studied the contact angle hysteresis occurrence in polymers and measured advancing and receding contact angles via the captive air bubble approach and the sessile liquid droplet method.

There have been a limited number of full scale theoretical studies investigating the water wettability for realistic and artificial surfaces.[21-24] Kimura and Maruyama[25] used two different potentials including Heinzinger- Spohr[26] and Zhu-Philpott[27] to model the interaction of water with a platinum surface, and found while the former cannot correctly predict the water contact angle, the latter provides accurate results. Ritos et al.[28] calculated and compared the wetting properties; such as the static, advancing and receding contact angle of graphite, silicon and an artificial superhydrophobic surface through molecular dynamics simulations. Peng et al.[29] used the Monte Carlo method and a mass density profile fitting approach to predict the water contact angle. Hirvi and Pakkanen[30] employed molecular modeling to evaluate the wetting tendency of two crystalline and amorphous polymer surfaces, polyethylene and polyvinyl chloride, and they found that by using the circular function to fit on the density profile, they can accurately predict the water contact angle.

In the following, we develop a pseudocontinuum model to probe the wetting phenomenon for different materials. In this model a surface interacts with water droplets via Lennard-Jones 9-3 potential. The effect of potential parameters on wettability of smooth surfaces is studied by carrying out molecular dynamics simulations. The water contact angle is calculated by the density fitting approach including the circular function. Simulations are performed in two distinct steps, the first of which is the preparation of a free water droplet, and the second of which involves the modeling of the droplet interacting with a smooth pseudocontinuum surface.

## 2. Computational Details

### 2.1. Simulation methodology

The simulations consisted of two steps. The first one involved generating a water droplet. Different droplets including 800, 1,200, 3,200, and 6,400 water molecules were generated by melting the ice structures at large NVT boxes including the periodic boundary conditions in all three directions. Equilibrium was achieved in a total time of 2 ns. To melt ice structures in a smooth way, several sub-steps including different temperatures were considered.

At the second step, the free droplet equilibrated at the previous step was transferred into another box containing two fixed L-J walls placed at the bottom and top of the box along the z direction. The periodic boundary conditions were only applied in the x and y directions. This step involved two sub-steps consisting of an equilibration time of 4-10 ns, depending on the size of the droplet and the strength of the wall, and was followed by 2 ns production time.

The rigid SPC/E[31] water model was used to simulate the interactions of water molecules. This model consists of an OH bond length of 1.0 Å, and a HOH angle of 109.47°. It also includes columbic (for both hydrogen and oxygen atoms) and 12-6 L-J (only for oxygen atoms) potential terms,

$$E_{Coul} = \frac{q_i q_j}{4\pi\epsilon_0 r_{ij}}, \qquad E_{LJ(12-6)} = 4\epsilon_{O-O}\left[\left(\frac{\sigma_{O-O}}{r_{ij}}\right)^{12} - \left(\frac{\sigma_{O-O}}{r_{ij}}\right)^6\right], \qquad (1)$$

here, $i$ and $j$ are atomic positions and $r_{ij}$ is the distance vector; $q_i$ and $\epsilon_0$ are the atomic charge ($q_O = -0.8476$ e and $q_H = +0.4238$ e) and vacuum permittivity, respectively. $\sigma_{O-O}$ and $\epsilon_{O-O}$ involve the oxygen L-J distance and energy parameters ($\sigma_{O-O} = 3.166$ Å and $\epsilon_{O-O} = 0.155$ kcal/mol), respectively. Also, surface potentials were assumed to be the 9-3 L-J,

$$E_{LJ(9-3)} = \epsilon_{surf}\left[\frac{2}{15}\left(\frac{\sigma_{surf}}{z}\right)^9 - \left(\frac{\sigma_{surf}}{z}\right)^3\right]. \qquad (2)$$

$z$, $\epsilon_{surf}$ and $\sigma_{surf}$ are the particle distance from the surface, the surface energy parameter, and the distance parameter, respectively (see Figure 4).

The Nóse-Hoover thermostat[32, 33] and the velocity Verlet scheme[34] with the time step of 1 fs were applied to control the temperature and atomic positions, accordingly. The cut-off radii of 10 Å and 12 Å were; respectively, considered for L-J and columbic interactions, and the latter interactions were described by a damped shifted model developed by Fennell and Gezelter[35]. The LAMMPS package[36] was employed to conduct molecular dynamics simulations.

### 2.2. Contact angle calculation

The mass density profile fitting procedure[22] was employed to obtain the water droplet contact angle. The method contains three paces[29]. Firstly, the two-dimensional (x-z and y-z) ensemble average density of liquid molecules, including a constant grid size is calculated. Then the average liquid-vapor interface density is determined and an isochoric curve is fitted with the interface points. The fit might include different functions such as polynomial, elliptical and circular functions. Since the density fluctuates tremendously close to the surface, generally a cutoff distance ($z_c$) is introduced and the points above this distance are only considered for fitting; however, the contact angle is measured by plotting a tangent line at another point ($z_0$).

In our study, density profiles were calculated at the production step of 2 ns. A grid size of 1 Å was used to compute the density every 500 time step resulting in a total number of 4000 profiles which were finally averaged to form the droplet mass density profile. Then, a circular function was fitted on the interface contour, and the data points with z less than $z_c$=6 Å were excluded. Next, the tangent line of the curve at $z_0$=2.5 Å was plotted, and the contact angle was calculated (see Figure 1).

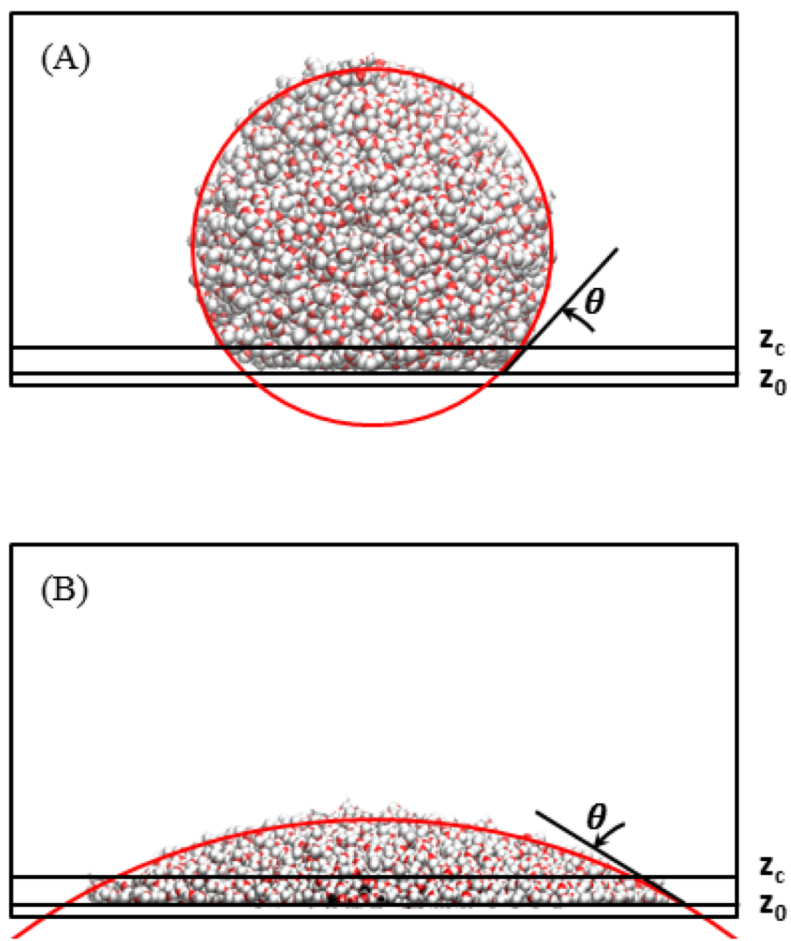

Figure 1. Water contact angle (θ) measurement by using circular fit on the water mass density profile. (A) the hydrophobic surface and (B) the hydrophilic surface. Color coding: red and white represent water oxygen and water hydrogen atoms, respectively.

## 3. Results and Discussion

### 3.1. Pseudocontinuum Surface Approximation

In this study, we used Lennard-Jones 9-3 potential to model the smooth wall. This model includes a pseudocontinuum wall interacting with fluid particles via 9-3 L-J potential which is obtained from the 12-6 L-J potential. Assume that a solid wall includes a lattice of 12-6 L-J atoms including a density $\rho$ (Figure 2(A)) and it interacts with the fluid particles through wall-fluid L-J parameters $\epsilon \cong \sqrt{\epsilon_w \cdot \epsilon_f}$ and $\sigma \cong (\sigma_w + \sigma_f)/2$ in which $\epsilon_w$, $\epsilon_f$, $\sigma_w$, and $\sigma_f$ are wall and fluid Lennard-Jones parameters. The total potential of the wall acting on a single particle of the fluid can be estimated as follows.

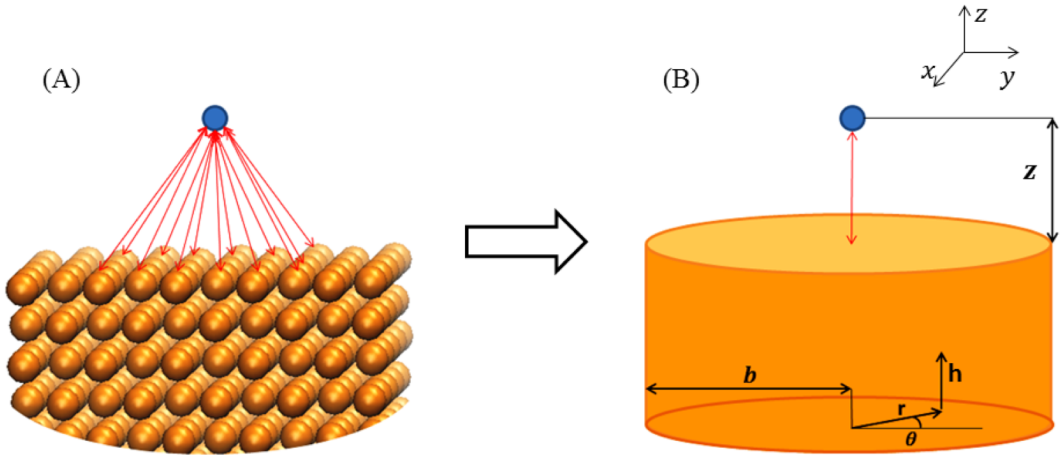

Figure 2. (A) a smooth structured wall containing 12-6 L-J particles interacts with a fluid particle. (B) a pseudocontinuum wall deals with the fluid particle via the 9-3 L-J potential.

$$E_w(z) = 4\epsilon\rho \int_0^{2\pi} d\theta \int_{-z-b}^{-z} dh \int_0^b r\, dr \left\{ \frac{\sigma^{12}}{(r^2+h^2)^6} - \frac{\sigma^6}{(r^2+h^2)^3} \right\}, \qquad (2)$$

now, consider b=∞ (Figure 3 illustrates finite b cases),

$$E_w(z) = 8\pi\epsilon\rho \int_{-\infty}^{-z} dh \int_0^{\infty} r\, dr \left\{ \frac{\sigma^{12}}{(r^2+h^2)^6} - \frac{\sigma^6}{(r^2+h^2)^3} \right\}, \qquad (3)$$

so,

$$E_w(z)=\epsilon_{surf}\left\{\frac{2\sigma^9}{15z^9}-\frac{\sigma^3}{z^3}\right\}, \qquad (4)$$

in which,

$$\epsilon_{surf}=\frac{2\pi\epsilon\rho\sigma^3}{3}. \qquad (5)$$

Therefore, by applying this model one might approximate all interactions of wall particles with a fluid particle by only a single wall-particle interaction depicted in Figure 2(B), and can save a considerable computational expense.

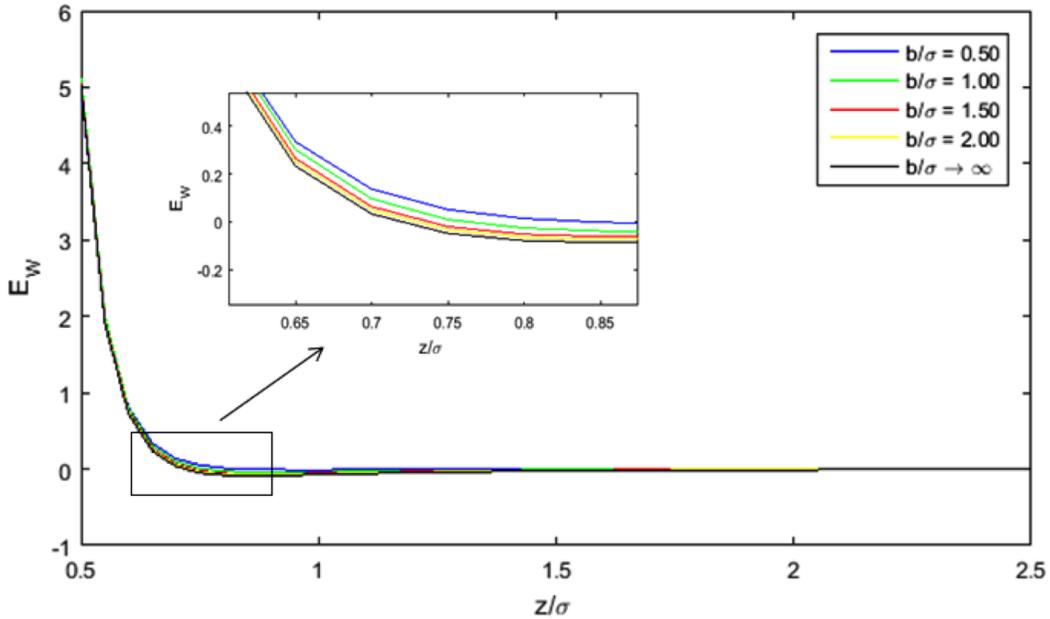

Figure 3. Wall-particle interaction energy as a function of the particle distance. (BCC structure)

Figure 3 compares the accuracy of the pseudocontinuum potential estimation for four different wall radii ($b/\sigma$ = 0.5, 1.0, 1.5, and 2.0) with respect to the exact result calculated from the summation of 12-6 L-J potential interactions. Figure 3 suggests that increasing the size of the surface leads to more accurate results. By computing the numerical integration, one may calculate the total amount of potential energy stemming out from the whole range of wall-particle interactions for a specific wall dimension. We calculated the area under the potential curve for BCC and FCC lattices with respect to b/σ dimensions (from 1.0 to 10.0). Next, the result is divided to that of the infinite wall (b/σ→∞) and plotted in Figure 4. The figure shows that for the surface dimensions larger than 2.0, the pseudocontinuum approximation can predict atomic interactions with accuracy of over 98 percent. Moreover, for the FCC lattice the potential converges faster to the infinite surface potential compared to the one from the BCC structure; however, both BCC and FCC potentials lead to reasonable results involving more than 99 percent accuracy at b = 3. Because of the nature of the droplet-surface interactions, for all simulations of this study, the wall dimensions were chosen to be at least 20 σ in the x and y directions.

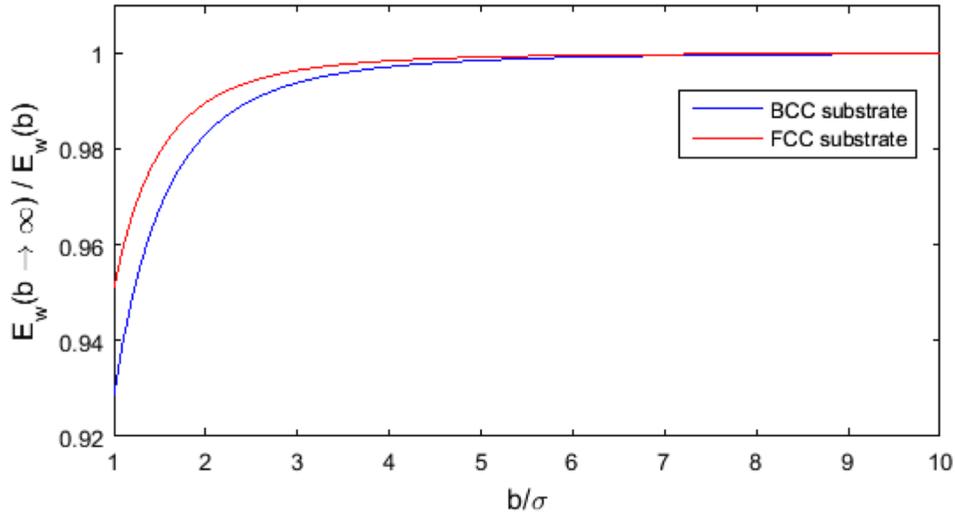

Figure 4. Accuracy ratio of pseudocontinuum approximation for BCC (blue line) and FCC (red line) lattices as function of the wall dimension.

### 3.2. Convergence of nanoscale to macroscale results

Using MD simulations, one can compute the contact angle from the mass density profile at nanoscale; although it might not be exactly comparable to the water droplet contact angle in the realistic conditions at the macroscopic level due to the massive size difference. To resolve this issue, we used modified Young-Laplace equation[14] which relates microscale contact angle to the macroscale one. This equation describes that at the equilibrium conditions where the free energy of the system is constant the relation between the force components can be written as,

$$\sigma_{sv} = \sigma_{sl} + \sigma_{lv}\cos\theta + \frac{\tau}{r}, \qquad (6)$$

$\sigma_{sv}$, $\sigma_{sl}$ and $\sigma_{lv}$ are the surface-vapor, surface-liquid and liquid-vapor tensions, respectively. $\tau$ is the line tension (free energy per unit length), r is the droplet-base radius and θ is the contact angle. The equation can be translated into,

$$\cos\theta = \cos\theta_\infty - \frac{\tau}{r\sigma_{lv}}, \qquad (7)$$

in which, $\theta_\infty$ is the contact angle of an infinitely large droplet and is defined as,

$$\theta_\infty = \cos^{-1}[(\sigma_{sv} - \sigma_{sl})/\sigma_{lv}]. \qquad (8)$$

Using Eqn. (7), one can compute the macroscale contact angle from nano-level calculations. The line tension would be obtained by carrying out different simulations including various droplet sizes and by plotting the contact angle (θ) vs $1/r$. Furthermore, the line tension is on the order of nanoscale where accurate contact angle measurements from lab experiments is demanding. Therefore, at this scale molecular dynamics method would be applied to remedy this difficulty and prepare rigorous contact angle computations.[37]

We modeled systems with different droplet sizes containing 800, 1200, 3200, and 6400 water molecules to predict the effect of the droplet size on accuracy of our calculations. Figure 5 illustrates variations in the contact angle of the droplets including three different surface energy strengths: $\epsilon_{surf}$ = 0.44, 0.84, and 1.64 kcal/mol. Figure 5 suggests that there is only a small deviation between the macroscopic contact angle and the one calculated from pseudocontinuum surface simulations containing 1200 water molecules; as a consequence, this method can result in both reasonable accuracy, and computational efficiency. Furthermore, one may notice the slope of the line connecting different points with a similar energy strength is always less than or equal to zero.

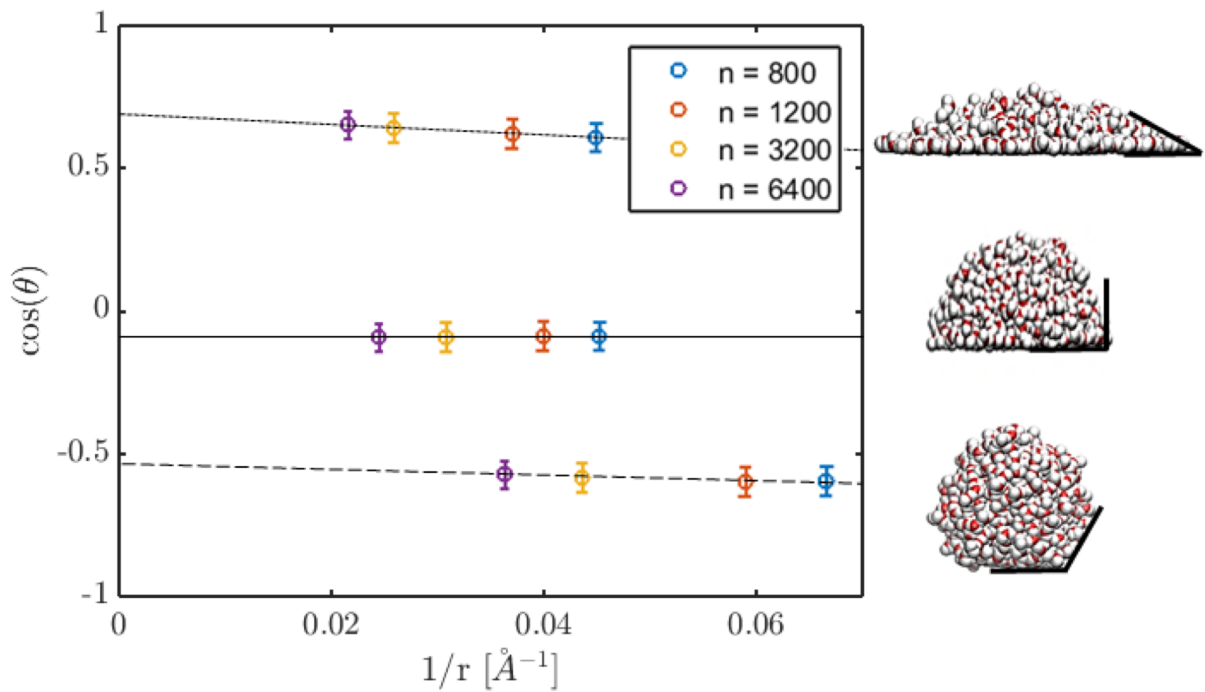

Figure 5. Water contact angle as a function of the inverse of base radius for wall strengths $\epsilon_{surf}=$ 0.44 kcal/mol (dashed line), 0.84 kcal/mol (solid line), and 1.64 kcal/mol (dotted line).

By using Equation (7), the ratio of $\tau/\sigma_{lv}$ was calculated for the three different walls including $\epsilon_{surf}=$ 0.44, 0.84, and 1.64 kcal/mol, respectively. Figure 6 suggests that the surface with the contact angle of ~ 95° results in the $\tau/\sigma_{lv}$ ratio of zero, while the hydrophobic and hydrophilic surfaces with contact angles ~ 125° and ~ 50° respectively, lead to higher ratios.

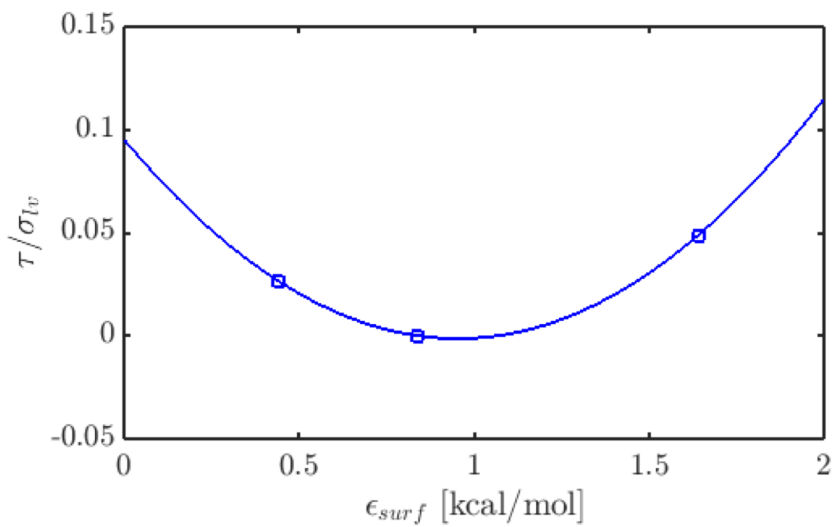

Figure 6. Calculated ratio of $\tau/\sigma_{lv}$ with respect to the wall strength $\epsilon_{surf}$.

### 3.3. Hydrophobicity and hydrophilicity

To determine the surface wettability, the water contact angle (θ) on different surfaces was studied. Variations of θ as a function of the L-J energy parameter ($\epsilon_{surf}$) at a constant temperature (T = 300 K) and a fixed water-surface distance parameter (σ = 3.12 Å) are shown in Figure 7. The result confirms that an increase in $\epsilon_{surf}$ intensifies the hydrophilic tendency of the surface as a result of stronger interactions between the surface and water molecules. At $\epsilon_{surf}$ = 0.94 kcal/mol, the surface is predicted to be neither hydrophobic nor hydrophilic (even-handed), while below and above this point the wall presents wetting and dewetting behaviors, respectively. This figure also suggests that there are two wetting regimes in one of which (θ > 14°) variations in $\epsilon_{surf}$ significantly affects the contact angle and is called the fast regime, whereas in the other one (the slow regime with θ < 14°) contact angle is more stable with respect to the wall strength.

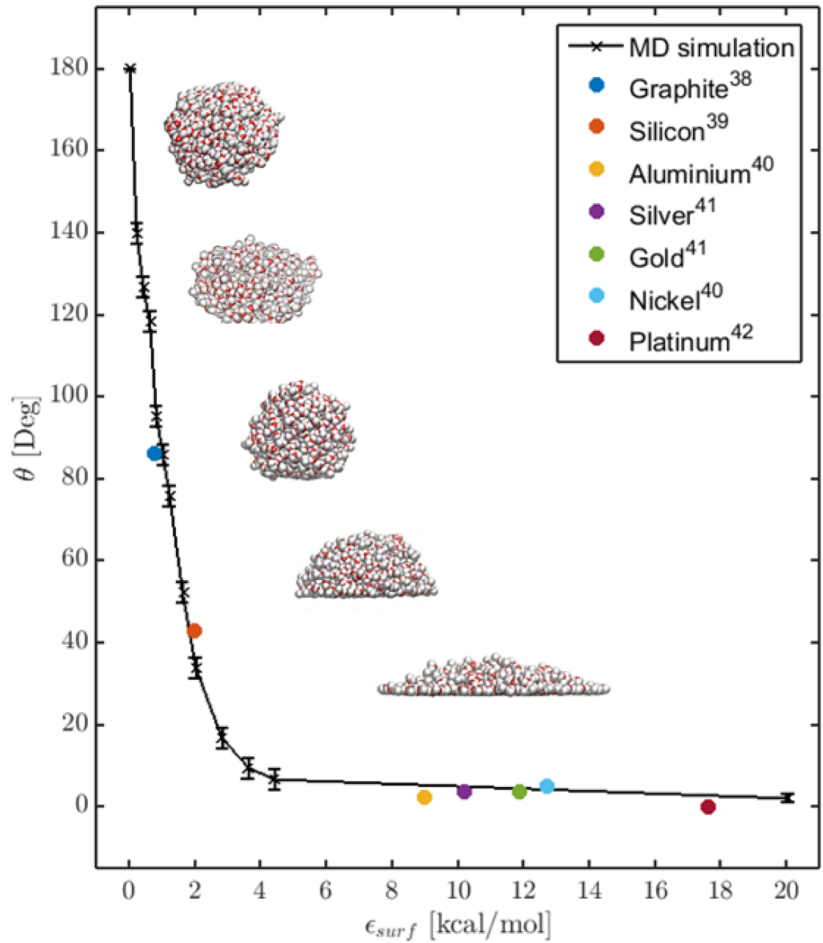

Figure 7. Water contact angle ($\theta$) as a function of the wall strength ($\epsilon_{surf}$). The droplet includes 1200 water molecules at T = 300 K and $\sigma$ = 3.12 Å. Color coding for the droplets consists of red and white for water oxygen and water hydrogen atoms, respectively.

Moreover, Figure 7 compares the contact angles from molecular simulations with the experimental ones for graphite, silicon, aluminum, silver, gold, Nickel, and Platinum structures consisting of $\epsilon_{surf}$ = 0.788, 1.988, 8.988, 10.218, 11.896, 12.706, and 17.620 kcal/mol, respectively. The $\epsilon_{surf}$ factor for experimental contact angles[38-42] was calculated by applying Eqn. (5) using the L-J parameters obtained from References 25 and 43. The MD results are in good agreement with the experimental data. There is a slight deviation about 7° for the silicon substrate with experimental angle[39] ($\theta = 43°$) from the calculated value ($\theta \cong 37°$) which can be answered by the fact that experimental measurement of Si contact angle is a demanding job due to the extensive reactivity of the Si surface with the air oxygen, creating a hydrophilic $SiO_2$ layer of 0.5 nm within one minute, increasing the contact angle of the silicon surface.[44]

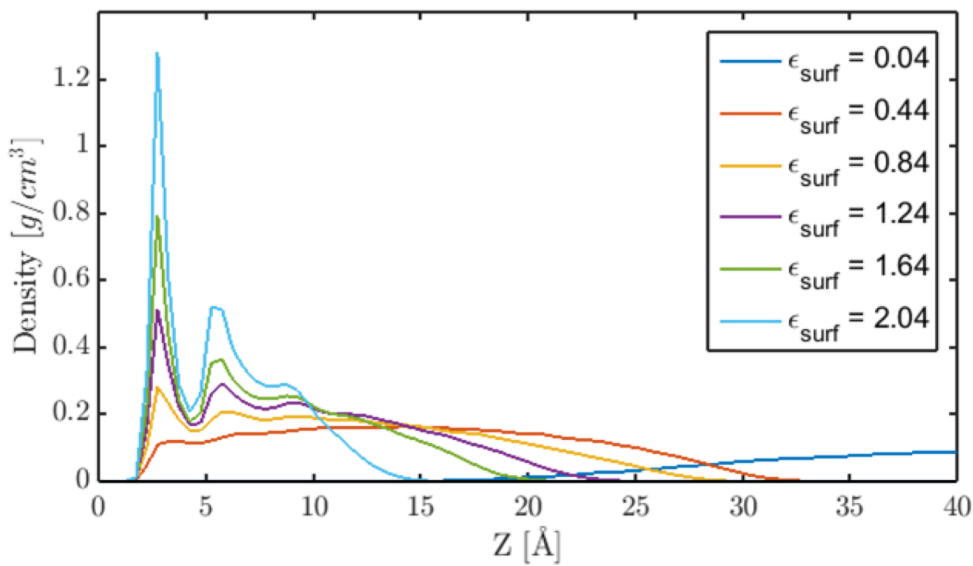

Figure 8. Water mass density with respect to the distance from the pseudocontinuum wall. $\epsilon_{surf}$ is in kcal/mol units.

Variation of water density with respect to the distance from the smooth wall in the z direction is depicted in Figure 8. There are several peaks on the density profiles presenting layered formation of water molecules in the z direction, which is explained by Maruyama et al.[45] The first and second peaks occur at $z = 2.7$ Å and $z = 5.75$ Å representing the first and second water layers. The number of water layers depends on the wall strength, and it decreases by the increase of surface wettability. For hydrophilic walls including $\epsilon_{surf} = 1.24$, 1.64 and 2.04 kcal/mol, the first peak is predicted to be the largest one which illustrates the prominent water configuration.

Moreover, for the super-hydrophobic surface ($\epsilon_{surf}$ = 0.04 kcal/mol) the maximal water density occurs at the middle of the box, and it seems that the surface repels water molecules. This phenomenon can be described by the fact that attractive forces between the liquid molecules is much stronger than the forces between the molecules and the wall, so liquid molecules colonize and form a spherical droplet in the center of the simulation box.

### 3.4. Effect of distance parameter and temperature

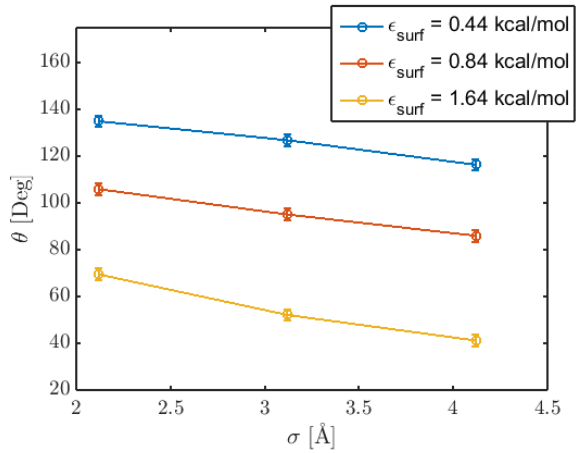 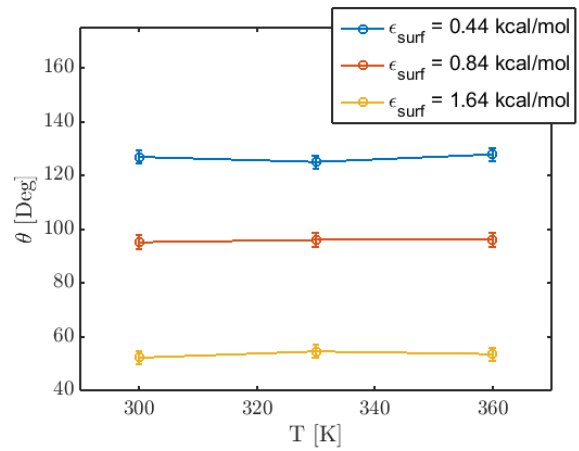

Figure 9. Surface contact angle (θ) as a function of the L-J 9-3 distance parameter (σ) with respect to. $\epsilon_{\text{surf}}$.

Figure 10. Surface contact angle (θ) as a function of the surrounding temperature (T) with respect to. $\epsilon_{surf}$.

Figures 9 and 10 depict variations of surface wettability with respect to the liquid-surface distance parameter and the temperature, respectively. Eqn. (4) consists of two terms including $\epsilon_{\text{surf}}$ and $\left\{\frac{2\sigma^9}{15z^9} - \frac{\sigma^3}{z^3}\right\}$. For plotting Figure 9, the latter term considered to depend on the distance parameter, while the first term was assumed to be independent of this parameter and was only determined by predefined $\epsilon_{\text{surf}}$. The distance factor σ plays a significant role in the surface wetting behavior and by the increase of this factor, the hydrophilic tendency of different surfaces consisting of hydrophobic, even-handed, and hydrophilic walls can be extended. Comparing the contact angle variation from Δθ = 18° $\epsilon_{surf}$ = 0.44 to Δθ = 28° for $\epsilon_{surf}$ = 1.64 with respect to σ describes that surfaces with higher energy strengths present larger fluctuations in the wetting tendency. As shown in Figure 10, surface contact angle is almost independent from the temperature ranging from T= 300 K to 360 K. This result is consistent with previous experimental studies.[17, 46]

## 4. Conclusions

We developed simple yet physically realistic molecular level theory to estimate contact angle/surface energy. Our approach will be useful in developing macroscopic design criteria for membrane/filtration system from molecular input. Molecular dynamics method was employed to investigate the wetting phenomenon of smooth surfaces. Two sets of simulations were carried out. First, a free liquid droplet was generated from the ice structure, and second the wall-droplet interaction was modeled. The former involved the fully periodic boundary conditions, while the latter involved a NVT box with periodicity in the x and y directions, and fixed walls in the z direction.

The pseudocontinuum wall model was derived from the structured wall model to substitute numerous 12-6 L-J atomic interactions between a fluid particle and wall atoms with only a single 9-3 L-J interaction to perform simulations at a lower computational cost. It was illustrated that the model can be leveraged to simulate different surfaces including limited dimensions. The results for BCC and FCC substrates were compared and it was predicted that FCC structures need smaller dimensions to produce accurate result consistent with that of the infinite wall.

To evaluate surface wettability, the droplet contact angel on the surface was probed. The angle was calculated from the ensemble average mass density and was measured by the circular fitting approach. Modified Young-Laplace equation (Eqn. (7)) was utilized to predict the accuracy of the simulations and to link the contact angle at nano- and macro-scales.

It was shown that the hybridized contact angle extensively depends on the energy strength dividing different surfaces into three categories including hydrophobic, hydrophilic, and evenhanded walls. This work also predicts two wetting regions according to the variation of contact angle with respect to the $\epsilon_{surf}$ parameter; one of them is more stable to the changes. The results of the pseudocontinuum modeling were compared to the experimental data for several substrates consisting of graphite, Si and FFC metallic surfaces, and it was illustrated that results are in substantial agreement. Only exception is for the Si experiment which involves a small inconsistency with our MD data. We believe this discrepancy can be ascribed to the inaccuracy of the experimental test due to the extensive reactivity of the Si surface to oxygen.

Next, the effects of the temperature and the liquid-surface distance factor on hydrophobicity were studied. Surface wetting was predicted to be independent of the temperature (for 300 <T<360 K), while it was found to be highly affected by the distance factor. The result also shows that the effect of this factor is more pronounced for the substrates including higher energy strengths.

In the future, we plan to investigate the wetting phenomenon by calculating the surface energy of the pseudocontinuum model. Using both of the contact angle and the surface energy computations leads us to better understanding of the phenomenon, less expensive study of different substrates, and more innovative design of the surfaces involving the desired wettability.

AUTHOR INFORMATION

**Corresponding Author**

* Email address: mj3a@andrew.cmu.edu

ACKNOWLEDGMENT

This technical effort was funded by Institute for Complex Engineered Systems (ICES), a member of the Pennsylvania Infrastructure Technology Alliance (PITA) program, at Carnegie Mellon University. Simulations were carried out on computers in Carnegie Mellon University and the Extreme Science and Engineering Discovery Environment (XSEDE) supported by National Science Foundation grant number ACI-1053575.